\documentclass[aps,prb,twocolumn,amsmath,amssymb,superscriptaddress,floatfix,scrartcl]{revtex4-1}

\usepackage{amssymb}
\usepackage{graphicx}
\usepackage{srcltx}
\usepackage{mathrsfs}
\usepackage[unicode=true,pdfusetitle,
 bookmarks=false,colorlinks=true,citecolor=blue,urlcolor=blue,linkcolor=red]{hyperref}

\begin{document}
\title{
Evolution operators in conformal field theories and conformal mappings:\\ 
the entanglement Hamiltonian, the sine-square deformation, and others}

\author{Xueda Wen}
\affiliation{
Institute for Condensed Matter Theory and
Department of Physics, University of Illinois
at Urbana-Champaign,
1110 West Green St, Urbana IL 61801
            }

\author{Shinsei Ryu}
\affiliation{
Institute for Condensed Matter Theory and
Department of Physics, University of Illinois
at Urbana-Champaign,
1110 West Green St, Urbana IL 61801
            }

\author{Andreas W.W. Ludwig} 
\affiliation{
Department of Physics,
University of California, Santa Barbara, CA 93106, USA}

\date{\today}

\begin{abstract}
By making use of conformal mapping, 
we construct various time-evolution operators in (1+1) dimensional conformal field theories (CFTs),
which take the form  $\int dx\, f(x) \mathcal{H}(x)$,
where $\mathcal{H}(x)$ is the Hamiltonian density of the CFT, 
and $f(x)$ is an envelope function. 
Examples of such deformed evolution operators 
include the entanglement Hamiltonian, and the so-called 
sine-square deformation of the CFT. 
Within our construction,
the spectrum and the (finite-size) scaling of the level spacing 
of the deformed evolution operator are known exactly.
Based on our construction, 
we also propose a regularized version of the sine-square deformation,
which,
in contrast to the original sine-square deformation, 
has the spectrum of the CFT defined on a spatial circle
of finite circumference $L$, and
for which the level spacing scales as  $1/L^2$,
once the circumference of the circle
and the regularization parameter are  suitably adjusted.
\end{abstract}


\maketitle


\section{Introduction}

Many classical statistical mechanical systems 
and quantum many-body systems at criticality enjoy conformal invariance -- 
invariance under  scale as well as special conformal transformations. 
Combined with translations and spatial rotations (or spacetime Lorentz boosts),
they are invariant under the conformal group. 
That critical systems are conformally invariant 
can be exploited to put some constraints on the operator 
content of the critical theory. 
Such constraints are most restrictive and powerful
in 2 or (1+1) dimensions, and in some cases can fully specify\cite{CardyOperatorContentNPB270-1986-p186}
the critical theory.
\footnote{
See also for example the reviews 
Refs.\ \onlinecite{DiFrancesco:1997nk, 2016arXiv160207982S}
and references therein.}

In this work, 
we consider various kinds of "deformations" of (1+1) dimensional
conformal field theories (CFTs). 
By "deformation" we mean the following. 
Let $\mathcal{H}(x)$ be the Hamiltonian density 
of a CFT where $x$ is the spatial coordinate.
Then, the ordinary time-evolution is generated by  
\begin{align}
H=\int\,dx\,\mathcal{H}(x).
\end{align}
We "deform" this Hamiltonian by introducing an "envelope function'' $f(x)$ as
\begin{align}
H[f]=\int dx\,f(x)\,\mathcal{H}(x).
\label{H deformed}
\end{align}
Similarly, suppose we have a lattice model, which is critical and
described by a CFT. Schematically, its Hamiltonian is given by
\begin{align}
H & =\sum_{i}h_{i,i+1}
\end{align}
where $h_{i,i+1}$ is the lattice analogue of the Hamiltonian density.
(We have restricted ourselves to the case of nearest-neighbor interactions,
and neglected for simplicity further neighbor interactions.
The lattice here can be periodic, infinite, or even open, but, we
are interested in the Hamiltonian density.) 
We "deform" this lattice Hamiltonian by introducing an "envelope"
function $f(x)$ as 
\begin{align}
H[f] & =\sum_{i}f\left(\frac{{x_{i}+x_{i+1}}}{2}\right)h_{i,i+1}.
\end{align}

There are various problems that fit into the above class of deformations. 
For example, 
let us consider the ground state $|\Psi\rangle$ of a CFT defined on 
infinite one-dimensional space, 
and then define the reduced density matrix $\rho_A$ associated with  a 
region $x\in (-R, R)$ by
$
\rho_A = \mathrm{Tr}_B\, |\Psi\rangle\langle \Psi|
$,
where the partial trace $\mathrm{Tr}_B$ is taken over the all degrees of freedom
associated with the region outside of the interval $(-R,R)$. 
Then, 
the entanglement Hamiltonian $H_E$,
defined by 
$
\rho_A = \exp (-H_{E}), 
$
is 
of the form \eqref{H deformed} 
where the envelope function is 
\begin{align}
f(x)=\frac{{R^{2}-x^{2}}}{2R},
\quad x \in (-R, R),
\end{align}
and $f(x)=0$ otherwise, 
\cite{Hislop:1981uh, Haag:1992hx, 2011JHEP...05..036C, 
CardyKITPTalk2015, Cho2016}
i.e., 
\begin{align}
	H_E=
	\int^R_{-R}dx\, \frac{R^{2}-x^{2}}{2R} \mathcal{H}(x).	
\end{align}

Another example is the so-called sine-square deformation (SSD)
of quantum many-body Hamiltonians in (1+1) dimensions.
\cite{Gendiar01102009,
Gendiar01022010,
2011PhRvB..83f0414H,
2011PhRvA..83e2118G,  
2011PhRvB..84k5116S,
2011PhRvB..84p5132M, 
2011JPhA...44y2001K,
2012JPhA...45k5003K, 
PhysRevB.86.041108,
PhysRevB.87.115128, 
2015JPhA...48R5208O,
2015MPLA...3050092T,
2015arXiv150400138I, 
2016arXiv160201190I}
In the SSD, one chooses the envelope function as 
\begin{align}
f(x)=\sin^{2}\frac{{\pi x}}{L},
\quad 
x \in (0, L),
\end{align}
and $f(x)=0$ otherwise. 
It was discovered that, for CFTs, 
the ground state of the SSD Hamiltonian is identical to
the ground state of the CFT defined on an infinite one-dimensional space.
This has  practical implications as the SSD 
Hamiltonian allows us to study 
the CFT in the thermodynamic limit  
by studying a
finite system of length  $L$
(in numerical simulations, say).

There are various other examples. 
For example, 
yet another context where such a  deformation has been discussed is the quantum energy inequalities. 
\cite{1997PhRvD..56.4922F}

Obviously, there are infinitely many ways to deform CFTs in this way.
As an attempt to find a systematic and controlled construction
of such deformations, we will make use of conformal mapping. 
Our construction can be described as follows: 
(i)
We start from a reference (1+1)-dimensional spacetime, 
parameterized by a complex coordinate which is denoted  in the following by $w$, 
and an evolution operator $\tilde{H}$.  
(ii)
We next pick a  suitable  conformal map which maps the reference 
space-time (coordinate $w$)  to the ``target'' spacetime, 
parameterized by a complex coordinate  which we denote in the following by $z$.
The conformal map maps the set of trajectories generated by $\tilde{H}$ 
(determined by the Killing vectors)
in the reference spacetime to some (potentially complicated)
trajectories in the complex $z$-plane. 
(iii) Finally, we transform $\tilde{H}$ and express it in terms of
the energy-momentum tensor on the complex $z$-plane. 
If we choose the reference evolution $\tilde{H}$ to be something simple, 
by construction, 
the spectrum of the deformed Hamiltonian is known exactly,
and so is its level spacing as a function of the  parameters on which the conformal
map depends (e.g.,
the system size). 
Put  differently, in our construction 
we deal with the set of envelope functions,
which we can ``undo'' by choosing a suitable  conformal map. 

The construction described above has been used, for example, 
to obtain the entanglement Hamiltonian in a number of cases.
\cite{PeschelTruong1987,CardyKITPTalk2015, Cho2016}
In this paper, by making use of conformal mapping, 
we describe various deformations of the CFT 
with various envelope functions, 
and also discuss the finite size scaling of their spectra.  
As a particular example, we obtain a ``regularized version'' of the SSD.
The regularized SSD is closed related to 
the entanglement Hamiltonian (defined for a finite interval),
in that
the entanglement Hamiltonian
and the regularized SSD
can be obtained from the same conformal mapping.
However,
the direction of the evolutions generated by them are 
orthogonal to each other.
(In the fluid dynamics language, the flows generated by 
these two evolution operators correspond to
the equipotential lines and the streamlines, respectively.)

As compared to the original SSD, 
the regularized SSD has the following properties: 
The spectrum of the regularized SSD Hamiltonian 
matches  the spectrum of a CFT with periodic boundary conditions (PBC).
However, 
the level spacing of  the regularized SSD Hamiltonian 
shows $(1/\mbox{length})^2$ scaling, as opposed to 
the familiar $(1/\mbox{length})$ scaling of a CFT with PBC. 
(To be more precise,
the length here means the length in the complex plane --
in the actual Hamiltonian, one needs to scale simultaneously both, the size of the
system and the regularization parameter.) 

On the contrary, 
the spectrum of the original SSD Hamiltonian
is known to possess a continuous spectrum (in the continuum limit and at criticality). 
For this reason, it is rather subtle to discuss the scaling of the finite size spectrum
of the SSD Hamiltonian on a lattice. 
Nevertheless,
it has been shown numerically that 
the spectrum of the SSD Hamiltonian on a finite lattice
shows $(1/\mbox{length})^2$ scaling.
(Once again, this should be contrasted with 
the ordinary $(1/\mbox{{length}})$ scaling of ordinary
CFT put on a finite cylinder. )
The regularized SSD does not have such subtle issues. 
The $(1/\mbox{length})^2$ scaling of
the regularized SSD may shed some light on 
the scaling of the original SSD on a lattice, 
by taking the  limit where the regularization parameter goes to zero.

By using the same idea, we also generated other deformations of CFTs.
For example, we obtain the "square root" deformation of CFTs, 
defined by the envelope function 
\begin{align}
	f(x) = \sqrt{(R-x)(R+x)}
	\label{envelope fn SRD}
\end{align}
for $x\in (-R, R)$ and 
$f(x)=0$ otherwise
(see Eq.\ \eqref{SRD cont}).
We will show that 
the level spacing of the deformed evolution operator
with the envelope function \eqref{envelope fn SRD}
does not depend on $R$.
This deformation was  previously discussed 
in the context of 
quantum information transport in quantum spin chains. 
In particular,  so-called `perfect state transfer' 
can be achieved in the XX model
with inhomogeneous 
nearest neighbor couplings
which 
are modulated according to the envelope function \eqref{envelope fn SRD}.
\cite{2004PhRvL..92r7902C, 2004PhRvL..93w0502A,2005PhRvA..71c2312C,2009arXiv0903.4274K}
\footnote{We thank Hosho Katsura for pointing out the connection
between the square-root deformation and the `perfect state transfer'.}


\section{Single vortex}
\label{Single vortex}

\begin{figure}
\includegraphics[scale=0.35]{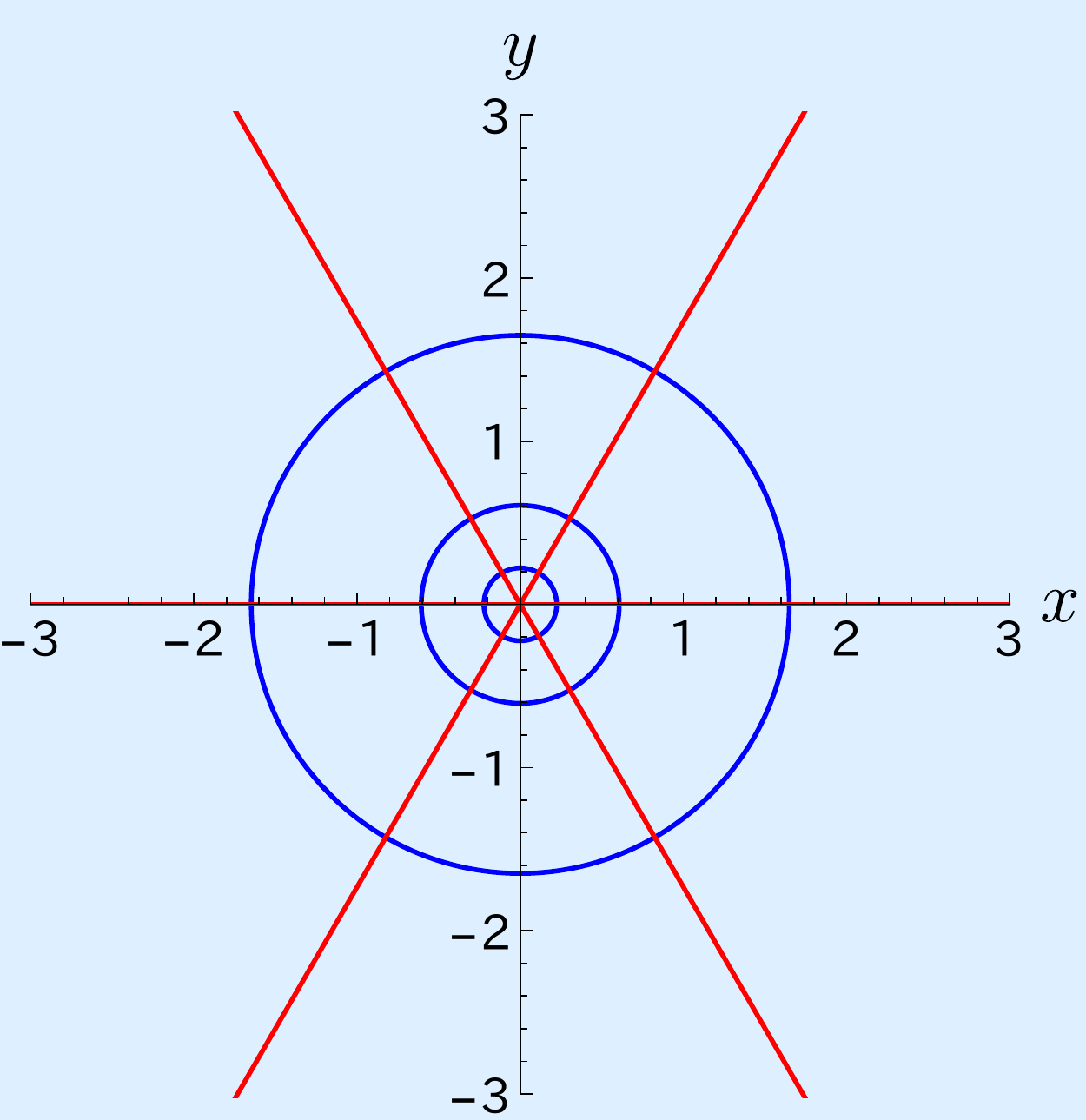} 
\includegraphics[scale=0.35]{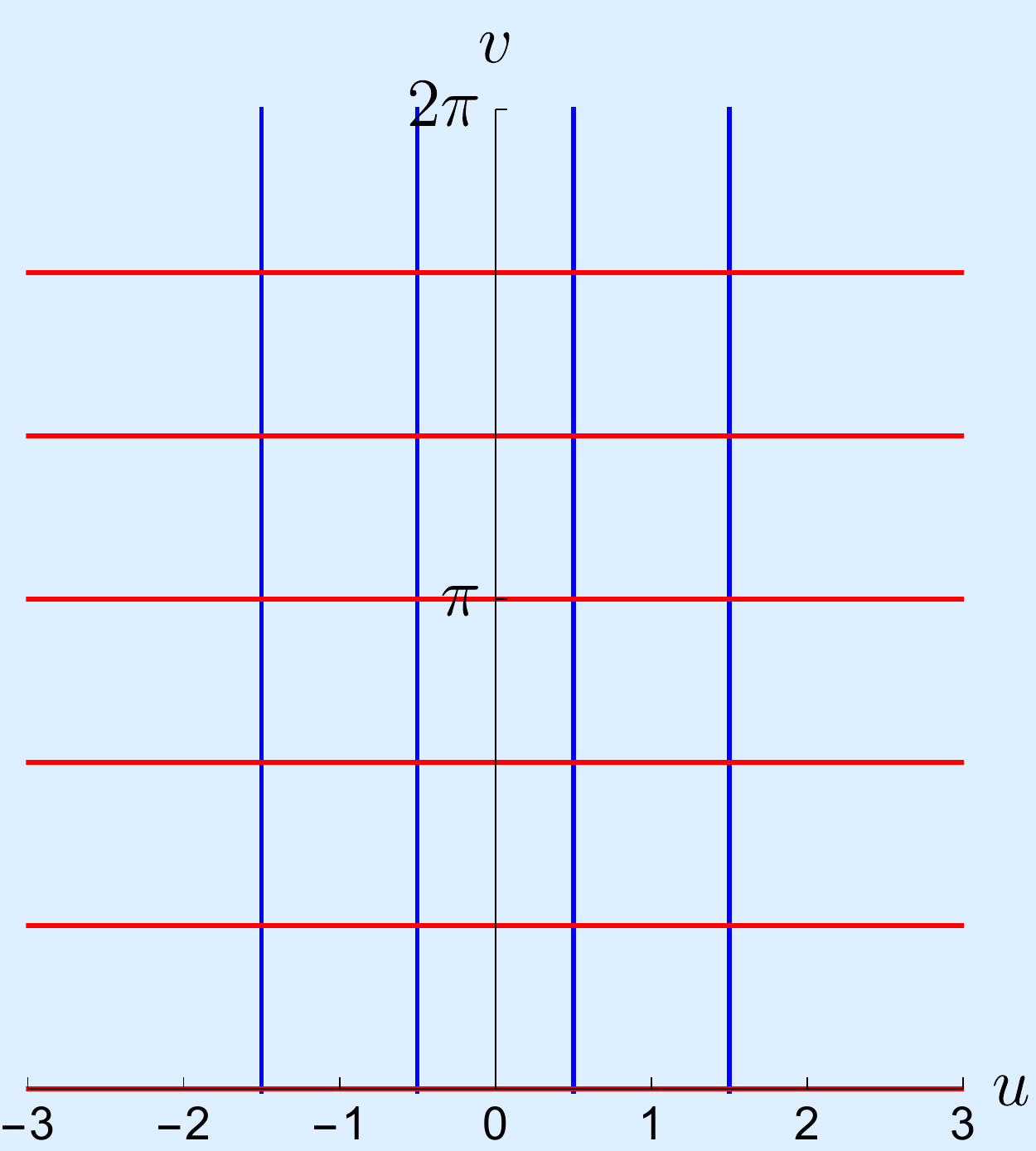}
\caption{Conformal map $w=\log z.$}
\end{figure}

We will consider conformal maps from 
the Euclidean spacetime to another (``target'') spacetime.
The ``target'' spacetime is parameterized by the complex coordinate 
$(z,\bar{z})$,
and we write the real and imaginary parts of $z$ as
\begin{align}
 z=x+iy.
\end{align}
The coordinate of the ``reference'' spacetime is denoted by 
$(w,\bar{w})$,
and we write the real and imaginary parts of $w$ as
\begin{align}
 w=u+iv. 
\end{align}

As a warm up, we start by illustrating our strategy by taking 
the well-known example of the 
radial and angular quantization of CFTs in  the complex plane. 
Consider the conformal mapping
\begin{align}
w(z)=\log z
\label{labelEqLogarithmicMap}
\end{align}
which maps the entire complex $z$-plane (``target'' spacetime) into an infinitely long cylinder ($w$-coordinates;
``reference spacetime'').
This conformal map also  transforms an annulus in the $z$-plane into a finite cylinder in the  $w$-plane. 
%
In the following, we will consider 
a `flow', or `time-evolution' along 
the $u$ or the $v$ direction in the ``reference'' spacetime.
We then consider the corresponding evolution in the ``target'' spacetime
by `mapping back' the evolution operator from the ``reference'' space time  into the $z$-plane. 

\vskip 1cm

\subsection{Radial flow ("radial quantization")}

The radial flow in the $z$-plane is mapped onto a "straight"
flow in the $w$-plane (flow along the  $u$-direction). This simple evolution
is generated by the evolution operator in the  $u$-direction,
\begin{align}
\tilde{H}=\int_{0}^{2\pi}dv\, \tilde{T}_{uu}(u_0,v),
\end{align}
where $\tilde{T}_{uu}$ is the $uu$ component of the stress (energy-momentum) tensor,
and we choose a particular ``time'' $u=u_0$ to define the evolution operator. 
This evolution operator is the Hamiltonian of a CFT on a 
circle of circumference $2\pi$. Since it is defined on a finite space, it 
has a discrete spectrum.

Mapping back to the $z$-plane, 
$\tilde{H}$ can be expressed in terms of the dilatation operator (a Virasoro generator) in the plane as
\begin{align}
 \frac{\tilde{H}}{2\pi} = L_{0}+\bar{{L}}_{0}-\frac{{c}}{24}
 \label{rad quantization}
\end{align}
where $c$ is the central charge and 
$L_{0}$ and $\bar{L}_0$ are given by
\begin{align}
L_{0}=\frac{{1}}{2\pi i}\oint dz\:z\:T(z),
\quad
\bar{L}_{0}=\frac{{-1}}{2\pi i}\oint d\bar{{z}}\:\bar{{z}}\:\bar{{T}}(\bar{{z}}).
\end{align}
The dilatation operator generates translations in the radial direction in the complex plane.
This is the well-known radial quantization.  

For later use, 
let us consider 
polar coordinates
$(r,\theta)$ in the $z$-plane defined by
\begin{align}
 z=r e^{i\theta}.
\end{align}
From the tensorial transformation law
of the energy-momentum tensor,
\begin{align}
&
T_{ij} = \tilde{T}_{\mu\nu} \frac{\partial u^{\mu}}{\partial x^i} \frac{\partial u^{\nu}}{\partial x^j},
\label{a4}
\end{align}
$\tilde{T}_{uu}$ can be expressed as
\begin{align}
 T_{rr}(r,\theta) = \tilde{T}_{uu} \frac{\partial u}{\partial r}\frac{\partial u}{\partial r}
 =
\frac{1}{r^2} \tilde{T}_{uu}(u,v) 
\end{align}
(up to the Schwartzian derivative term).
Hence, $\tilde{H}$ can also be expressed as 
\begin{align}
\tilde{H}=
r^2_0
\int_{0}^{2\pi}dv\, T_{rr}(r_0, \theta)
=
r^2_0
\int_{0}^{2\pi}d\theta\, T_{rr}(r_0, \theta),
\end{align}
where $r_0$ is defined by $u_0=\log r_0$.

Since the circumference of the circle in the $z$-plane is 
$
	L := 2\pi r_0,
$
it is natural to introduce 
\begin{align}
	\theta = \frac{2\pi}{L} s,
	\quad
	s\in [0, L],
	\quad 
	L = 2\pi r_0. 
\end{align}
Then, $\tilde{H}$ can be written as
\begin{align}
	\tilde{H} = \frac{L}{2\pi} \int^{L}_0 ds\, 
	T_{rr}\left(\frac{L}{2\pi},\frac{2\pi s}{L}\right). 	
\end{align}
Comparing with Eq.\ \eqref{rad quantization}, 
the spectrum of the operator 
\begin{align}
\int^{L}_0 ds\, T_{rr}\left(\frac{L}{2\pi}, \frac{2\pi s}{L}\right)
\label{reference ham}
\end{align}
scales as $1/L$, and hence
this operator can be considered\cite{CardyFSSJPhysA17-1984-L385}  as 
the Hamiltonian of a CFT defined on a circle of circumference $L$.
\footnote{See e.g. also Ref.\ \onlinecite{Cardy:1989da} for a review. }

\subsection{Angular flow ("Rindler Hamiltonian")}

The conformal transformation in Eq. (\ref{labelEqLogarithmicMap})
also maps the  angular flow in the $z$-plane  into the "straight"
flow along the $v$-direction  in the $w$-plane.
This simple time-evolution
is generated by the evolution operator in $v$ 
direction\footnote{Technically, 
$\tilde{H}=\int_{u_{-}}^{u_{+}}du\: \tilde{T}_{vv}$, 
where
 $u_{\pm} \equiv\ln( R_{\pm}/a)$, and
 $ R_-/a < |z| < R_+/a$ describes an annulus in the $z$-plane with $a$ a short-distance cutoff; we are ultimately
interested in the limit $R_{-}/a \to 1$ and $R_{+}/a\to \infty$, corresponding to $u_{\pm}\to \pm \infty$.}
\begin{align}
\tilde{H}=\int_{-\infty}^{+\infty}du\: \tilde{T}_{vv}(u,v).
\label{labelEqAngularFlowHamiltonian}
\end{align}
We next transform this evolution operator back into the $z$-plane. 
Using the tensorial transformation law \eqref{a4}, 
the energy-momentum tensor on the cylinder and 
in the plane are related by 
\begin{align}
 T_{yy} &=
\frac{1}{(x^2+y^2)^2}
\left[
y^2 \tilde{T}_{uu}
 +
(xy)^2 (\tilde{T}_{uv}+\tilde{T}_{vu})
 +
 x^2 \tilde{T}_{vv}
\right].
\end{align}
In particular, when $y=0$ we have the relationship
$
 T_{yy} = x^{-2} \tilde{T}_{vv}, 
 $
and hence when expressed in the $z$-plane,
$\tilde{H}$ 
 is given by
\begin{align}
 \tilde{H} &= \int du \tilde{T}_{vv}
=
\int^{\infty}_0 dx\, x T_{yy}. 
\label{labelEqRindlerHamiltonian}
\end{align}
This is the  Hamiltonian in Rindler spacetime. 
The Rindler Hamiltonian
Eq. (\ref{labelEqRindlerHamiltonian})
 generates the angular flow (the Lorentz-boost in Minkowski signature), and corresponds to
``angular quantization'' in the $z$-plane.
It is also nothing but the entanglement Hamiltonian 
of the reduced density matrix
associated to the semi-infinite interval $x\in (0, \infty)$
(where we consider the ground state $|\Psi\rangle$ of a CFT defined on 
 infinite one-dimensional space, 
and then take the partial trace over all degrees of freedom
for $x\in(-\infty, 0)$.) - Since the entanglement or  Rindler Hamiltonian (or Lorentz-boost) in
 Eq. (\ref{labelEqRindlerHamiltonian}) 
is equal to the Hamiltonian Eq. (\ref{labelEqAngularFlowHamiltonian}) of the CFT
defined on an infinite space, its spectrum is continuous.\footnote{Note however that the entanglement Hamiltonian
of a {\it gapped} theory, in contrast to the one of the gapped CFT) theory discussed here, has a discrete spectrum,
as was shown in Ref. \onlinecite{Cho2016}.}

%
%
%

%
%

\section{Vortex-anti-vortex pair}

\begin{table}
\includegraphics[scale=0.35]{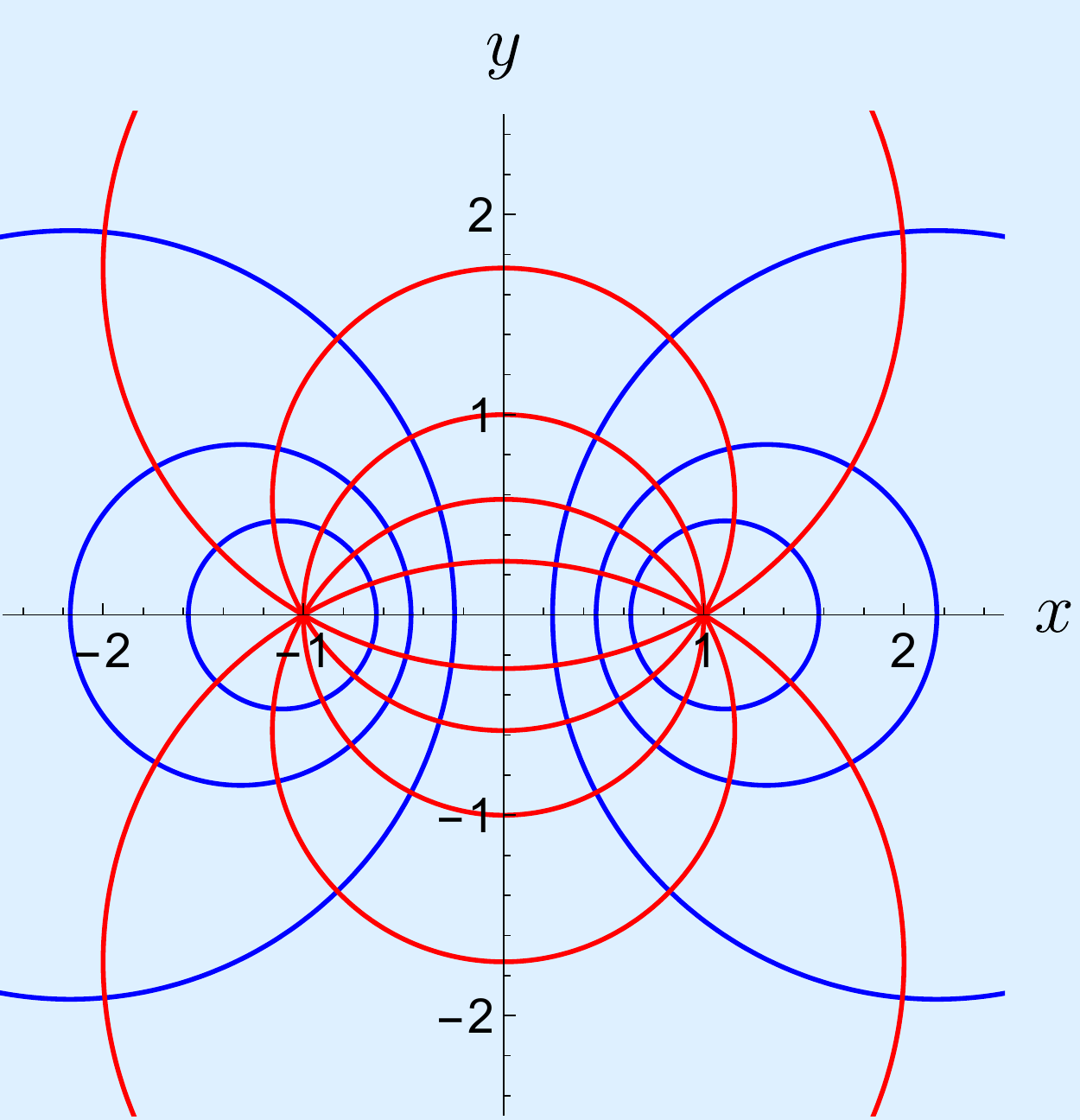}
\includegraphics[scale=0.35]{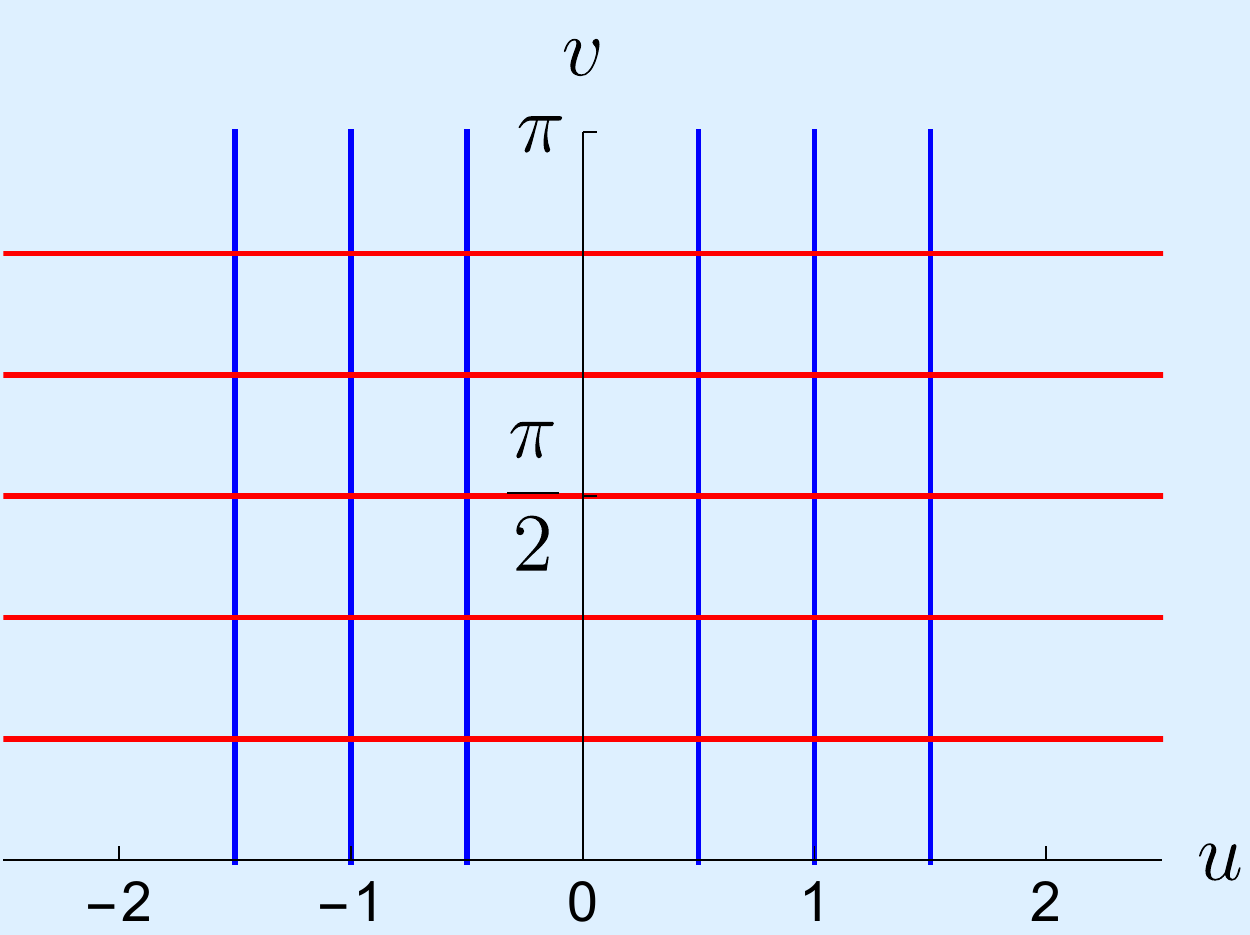}
\caption{Conformal map $w=\log(z+1)/(z-1).$}
\label{labelTABLEI}
\end{table}

In this section, we consider the conformal map
\begin{align}
w(z)=\log(z+R)-\log(z-R),
\label{double vtx}
\end{align}
with inverse $z(w)=$ $R \coth (w/2)$.
This maps the entire complex $z$-plane into an infinitely long  cylinder. The coordinate $v$ in the ``reference''
spacetime (coordinates $w$) is periodic with period $=2\pi$. 
%
This conformal map can be thought of as describing 
the complex potential of a flow which consists of 
a source with unit strength  located at $z=-R$,
represented by the complex potential $\log(z+R)$,
and 
a sink with the same strength located at $z=+R$.
%

Taking the limit $R/z \to 0$,
the vortex-anti-vortex pair reduces to a dipole, 
\begin{align}
\log\frac{{z+R}}{z-R} 
 & \sim\frac{{2R}}{z}.
 \label{dipolar limit}
\end{align}
It is also convenient to consider 
a pair of a vortex of strength $k$ and an anti-vortex of strength $-k$.
Then, by letting $R\to 0$ and $k\to\infty$ such that $2kR$ is finite, 
\begin{align}
w(z)=
k\log\frac{{z+R}}{z-R} 
 & \sim\frac{{2Rk}}{z}.
\label{labelEqDipole}
\end{align}
This is the complex potential due to a 
dipole, i.e., the
combination of a source and sink of equal strengths separated by a
very small distance. 
The quantity $2kR$ is the dipole moment. - Note that the inverse of the left equation in 
Eq. (\ref{labelEqDipole}) reads $z(w)=R\coth(w/2k)$,
which shows that the period of $v$ in the ``reference'' spacetime tends to
infinity in the dipolar limit, $k\to \infty$. This is consistent with the fact that `dipolar map' $w=1/z$ maps the entire complex plane into itself.
(See also Section \ref{labelSectionDipolarMap}.)

\subsection{$v$-evolution: Entanglement Hamiltonian}

As in the simple exercise we did in Sec.\ \ref{Single vortex}, 
we now consider two kinds of time-evolutions associated to the 
conformal map \eqref{double vtx}.

Let us first take $v$ as time, 
and consider the evolution operator 
at\footnote{The $v$-``time''-evolution maps the Hamiltonian defined at $v=0$ on the space (Cauchy-Surface)
$u\in$ $(-\infty,-R)\cup (+R, +\infty)$ to the Hamiltonian defined at $v=\pi$ on its complement, i.e. on  the space (Cauchy-Surface)
$u \in$ $(-R, +R)$.}
 $v=v_0=\pi$,
\begin{align}
\tilde{H}=\int_{-\log[(2R-\epsilon)/\epsilon)}^{+\log[(2R-\epsilon)/\epsilon]}du\, \  \tilde{T}_{vv}(u,v_0=\pi). 
\label{v-evolv double vtx}
\end{align}
Here, we have cut off the integral over $u$ by 
introducing\footnote{This cutoff is motivated as follows: when $y=0$,
and $-R<x<R$, we have 
$ 
e^{u+iv_0}=({x+R})/({x-R})
$
with $u=$ real
when $v_0=\pi$.
Then, taking $x=R-\epsilon$ or $x=-R+\epsilon$, we obtain
$u=\pm\log(2R-\epsilon/\epsilon)$
}
an UV cutoff $\epsilon>0$ in position space $x$ (so that $-R+\epsilon < x < +R-\epsilon$).

The (time-)evolution operator in \eqref{v-evolv double vtx}  generates the evolution along the constant $u$-trajectories. 
In the fluid dynamic terminology, these are equipotential lines and are given in the $z$-plane  by
\begin{align}
[x\pm R\,\coth(u)]^{2}+y^{2}=\frac{{R^{2}}}{\sinh^{2}(u)}.
\label{const u}
\end{align}
Thus, the constant $u$ trajectories are, 
for different values of $u$, 
circles having centers at $(x,y)=(\pm R\coth(u),0)$ and radii equal to $R/|\sinh(u)|$.
(Compare TABLE \ref{labelTABLEI}.)

The evolution operator \eqref{v-evolv double vtx}
can be now mapped into the $z$-plane. 
Focusing on $y=0$, 
$\tilde{T}_{vv}$ is transformed as 
\begin{align}
 T_{yy} &= \tilde{T}_{vv} \left(\frac{\partial v}{\partial y}\right)^2 
 = 
\tilde{T}_{vv}\left[  \frac{2R}{(x-R)(x+R)}\right]^2,
\end{align}
and hence $\tilde{H}$, when mapped into the $z$-plane, reads
\begin{align}
\tilde{H}=\int_{-R+\epsilon}^{+R-\epsilon}dx\frac{{(x-R)(x+R)}}{2R}T_{yy}. 
\end{align}
This is the entanglement Hamiltonian
obtained from a CFT defined on an infinite line,
after tracing out degrees of freedom living outside of the finite interval $x\in [-R, R]$.\cite{Hislop:1981uh, Haag:1992hx, 2011JHEP...05..036C, 
CardyKITPTalk2015, Cho2016}

%
%
%

\subsection{$u$-evolution: ``Regularized'' SSD}

Let us now move on to 
consider the evolution operator along the $u$-direction in  $w$-space which is given by 
\begin{align}
	\tilde{H} =\int^{\pi}_0 dv\, \tilde{T}_{uu}(u_0,v)
	\label{u evolv rSSD}
\end{align}
where we fix $u=u_{0}$. 
The constant-$v$ trajectories under this  evolution,
which are `streamlines' in the fluid dynamics language, 
are given in the $z$-plane  by
\begin{align}
x^{2}+[y+R\cot(v)]^{2}=\frac{{R^{2}}}{\sin^{2}(v)}.
\end{align}
For different values of $v$, these are circles with centers at 
$(0,-R\cot(v))$,
and radii $R/|\sin(v)|$. These circles pass through $(\pm R,0)$. - See TABLE \ref{labelTABLEI}.

Turning now  to the constant-$u$ ``time''-slices (the Cauchy surfaces 
for  the current choice of  ``time''-evolution), 
we see from Eq.\ \eqref{const u} that
their  $(x,y)$-coordinates  in the $z$-plane satisfy, for a fixed $u=u_0$, 
\begin{align}
\left(x+\frac{{\cosh u_{0}}}{\sinh u_{0}}R\right)^{2}+y^{2}=\frac{{R^{2}}}{(\sinh u_{0})^{2}}.
\end{align}
These are  circles of radius $r_0$ and circumference $L$, where
\begin{align}
r_0:=\frac{{R}}{\sinh u_{0}},
\quad  {\rm and} \ \ 
	L=2\pi r_0.
	\label{param}
\end{align}
The evolution operator \eqref{u evolv rSSD}, defined at $u=u_0$, 
acts on quantum states defined on the circle.

Making use of
the transformation law of the energy-momentum tensor, 
\begin{align}
	T_{rr}= 
	\tilde{T}_{uu} \left(\frac{\partial u}{\partial r}\right)^2
	=
	\tilde{T}_{uu} \frac{1}{r_0^2}
	\left(\frac{\sinh u_0}{\cos\theta + \cosh u_0}  \right)^2, 
\end{align}
$\tilde{H}$ can be mapped into the $z$-plane and can be written as 
\begin{align}
	\tilde{H} 
	&=
	r_0^2
        \int^{2\pi}_0 d\theta\,  
	\frac{\cos\theta +\cosh u_0}{\sinh u_0}\, 
	T_{rr}(r,\theta). 
\end{align}
By further introducing 
\begin{align}
	\theta =\frac{2\pi}{L} s, 
	\quad
	s\in [0,L], 
\end{align}
$\tilde{H}$ is written as
\begin{align}
	\tilde{H} 
	&= 
	\frac{L}{2\pi}
	\frac{1}{\sinh u_0} 
	\int^{L}_0 ds\,
	\nonumber \\
	&\quad 
	\times 
	\left(\cos \frac{2\pi s}{L} + \cosh u_0\right) 
	T_{rr}\left(r=\frac{L}{2\pi},\theta= \frac{2\pi s}{L}\right). 
\end{align}
From the discussion
in the paragraph containing
Eq.\ \eqref{reference ham}, 
the operator 
$
	\int^{L}_0 ds\, T_{rr}\left(L/2\pi, 2\pi s/L\right) 
$
is the Hamiltonian of a CFT defined on a circle of circumference $L$.
Thus, the part of $\tilde{H}$
which we call 
$H_{\mathrm{rSSD}}$, defined by
\begin{align}
	H_{\mathrm{rSSD}}
	=
	\int^{L}_0 ds 
	\left(\cos \frac{2\pi s}{L} + \cosh u_0\right) 
	T_{rr}\left(\frac{L}{2\pi}, \frac{2\pi s}{L}\right), \label{RegularizedSSDHamiltonian}
\end{align}
is the "deformed Hamiltonian" with the envelope function
$
\left (\cos (2\pi s/L) + \cosh u_0 \right ) 
$.
Because of the presence of the term $\cosh u_0$,
this deformation is different from the ordinary SSD,
and can be regarded as a "regularized" version of the SSD. 
The limit $u_0\to 0$ corresponds to the ordinary SSD, 
which for fixed $L$, or equivalently for fixed $r_0$, corresponds to $R\to 0$. (See
Eq. (\ref{param}).) As discussed 
in the paragraph containing Eq.\ (\ref{labelEqDipole})
this is the dipolar limit. 

By construction, 
when $u_{0}\neq$0,
the spectrum of the $u$-evolution operator is
the spectrum of a CFT  defined on a finite circle (i.e., with PBC),
which is discrete. 
This should be contrasted with the ordinary SSD, 
for which the spectrum of the evolution operator is a  continuum.

\paragraph{Finite Size Scaling}

We now turn to the finite-size scaling of the spectrum of the
regularized SSD evolution operator, Eq. (\ref{RegularizedSSDHamiltonian}).
(Once again, in the SSD limit $u_0 \to 0$,
the spectrum is continuous, and hence there is no finite size scaling to 
discuss.) 
We can in principle discuss the following two kinds of finite-size scaling behaviors.

First, we fix $u_{0}$ and change the distance $R$  between the two monopoles
which controls the (spatial)  size of the system. 
Since $\tilde{H}$ has 
a  level spacing of order one,
recalling \eqref{param},
the level spacing of $H_{\mathrm{rSSD}}= (2\pi \sinh u_0/L) \tilde{H}$
scales as
 \begin{align}
\sim \frac{\sinh u_0}{L}
={1\over 2 \pi} 
\frac{(\sinh u_0)^2}{R}
\sim 
\frac{1}{R}.
\end{align}
Since $R$ is proportional to $L$ when $u_0$ is fixed, 
this means  $1/L$ scaling. 

On the other hand, 
we can fix $R$ and change $u_0$
which, due to \eqref{param},
 also controls the size  $L$ of the system. 
In this case, the level spacing scales as 
\begin{align}
\sim \frac{\sinh u_0}{L}
=
{1\over 2 \pi} 
\frac{(\sinh u_0)^2}{R}
\sim
\frac{1}{L^2}.
\end{align}
This should be contrasted with  
the regular $1/L$ scaling of CFTs put on a finite spatial circle of 
circumference $L$. 
It should also be  noted that 
for the original SSD, 
 previous numerical studies 
reported $1/L^2$ scaling.
\cite{Gendiar01022010,
PhysRevB.87.115128}
The $1/L^2$ scaling of the regularized SSD may be related to this observation. 
Finally, as we will discuss momentarily, 
for the original SSD the spectrum consists of a  continuum, when 
we consider the continuum field theory (CFT) formulation of the system, 
and hence there is no finite size scaling  to discuss.

\paragraph{The Dipolar Limit}

Let us now consider the dipolar limit $R\to 0$. 
In the dipolar limit, the spacetime cylinder 
in  $w$-space
shrinks
as it is bounded in the $u$-direction by the cut off $\pm\log [(2R-\epsilon)/\epsilon]$.
- See 
Eq. (\ref{v-evolv double vtx}). (Note that in the  section containing
Eq. (\ref{v-evolv double vtx}),
  we discussed the entanglement Hamiltonian
where $u$ represented the ``spatial'' coordinate, and $v$ the (imaginary time) ``temporal'' coordinate.
In contrast, in the present section, we have chosen the  $u$-direction as our (imaginary time)
 ``temporal'', and the $v$-direction as our "spatial" coordinate.)
Thus, in the limit $R\to\epsilon$, the (``temporal'')  $u$-direction shrinks to zero.
Hence,
the "modular parameter" of the CFT,
which depends on the aspect ratio of space and (imaginary)  time directions, 
is given by 
\begin{align}
\mbox{(total space length)}/\mbox{(total time length)} 
\to \infty.
\end{align}
This can be interpreted as achieving the infinite size limit,
as noted by Ishibashi and Tada.
\cite{2015arXiv150400138I,2016arXiv160201190I}

Before closing this section, 
let us discuss the special case $u=0$.
This means that we consider the evolution in the $z$-plane right on the imaginary axis, $x=0$,
rather than on  a finite circle that we considered when $u\neq 0$.  
Hence, we consider the evolution (flow)
that 
brings the infinite line to a point (eventually, at asymptotically long times).  - See TABLE \ref{labelTABLEI}.
The evolution operator 
$
\tilde{H} =\int dv \tilde{T}_{uu}(u=0,v)
$
is mapped into 
\begin{align}
\tilde{H}=\int_{-\infty}^{+\infty}\,dy\, \ \frac{y^{2}+R^{2}}{2R} T_{xx}.
\end{align}
By construction, 
this Hamiltonian $\tilde{H}$ has a spectrum which is described by a CFT with PBC,
although the system is defined on infinite one-dimensional space. 
%
%
%
In the dipolar limit, $R \ll y$, this yields 
\begin{align}
\tilde{H}=\frac{1}{2R}\int\,dy\frac{{1+(R/y)^{2}}}{y^{-2}}
T_{xx}
\sim\frac{{1}}{2R}\int\,dy\,y^{2} T_{xx}.
\end{align}
This evolution
operator can be considered as derived from the decompactification
limit of the SSD Hamiltonian:
$
\int_{0}^{L}\,dx\,\sin^{2}\frac{{\pi x}}{L}\mathcal{{H}\sim\int}dx\,x^{2}\mathcal{{H}}.
$
Observe that
while before taking the dipolar limit, 
the system is defined on the whole imaginary axis;
the limit $R\to0$ "cuts" the imaginary axis into two halves, $y>0$ and $y<0$. 

\section{Dipolar map}
\label{labelSectionDipolarMap}

\begin{figure}
\includegraphics[scale=0.35]{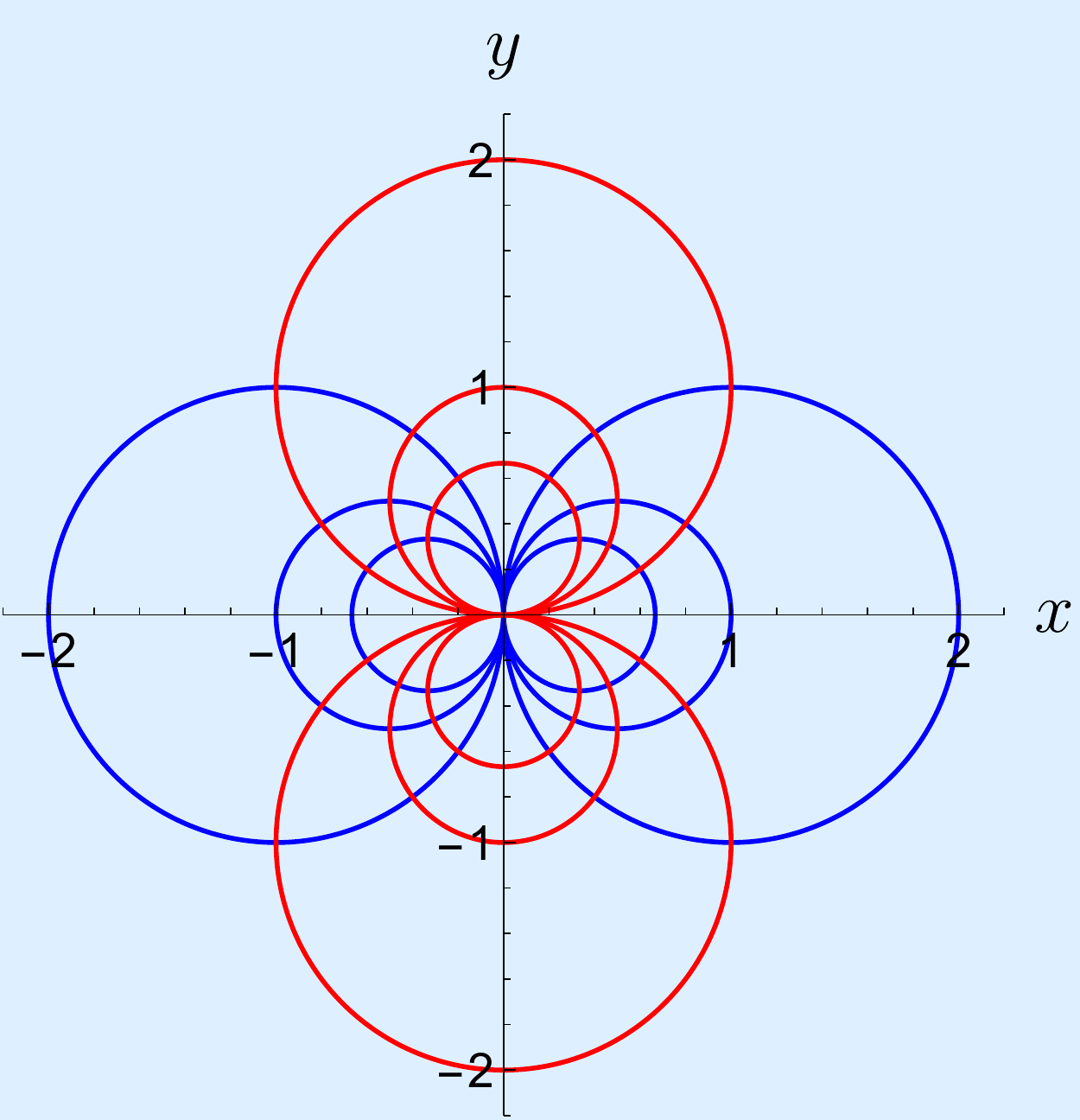}
\includegraphics[scale=0.35]{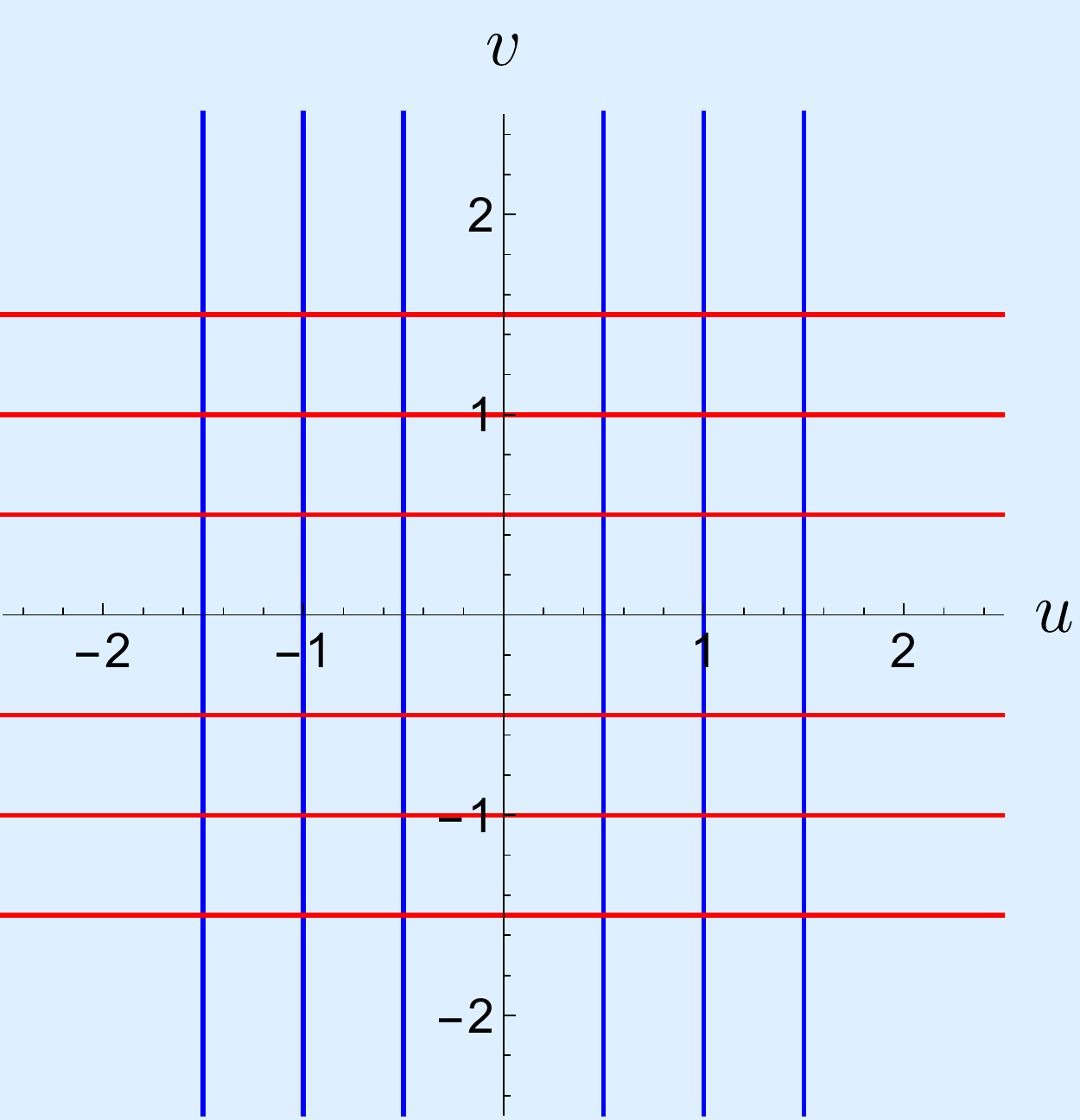}

\caption{Conformal map $w=1/z$.}
\end{figure}

%
%
%
%

The conformal map 
\begin{align}
	w= \frac{1}{z}
\label{labelEqDipolarFlow}
\end{align}
describes the dipolar flow, which maps the entire complex $z$-plane into the entire complex $w$-plane.
As described  in the paragraph containing Eq.\ (\ref{labelEqDipole}),
the corresponding flow can be obtained from a pair of a vortex and an anti-vortex,
by taking the limit where their separation goes to zero. 
Here, we directly deal with the dipolar flow without taking the limit. 

We will focus on the evolution in $u$-direction. Furthermore, we
use the parametrization $u=1/(2r_0)$. 
For $u=1/(2r_0)$, Eq.(\ref{labelEqDipolarFlow})
implies
\begin{align}
\frac{{1}}{2r_0}=\frac{{x}}{x^{2}+y^{2}}\Rightarrow(x-r_0)^{2}+y^{2}
=r^{2}_0.
\end{align}
Thus, the constant $u$ trajectories are 
circles of radius $r_0$ centered at $(r_0, 0)$. 

We consider the evolution operator in  $w$-space given by 
\begin{align}
	\tilde{H} =\int_{-\infty}^{+\infty} dv\, \  \tilde{T}_{uu}(u_0,v),
\label{labelEquEvolutionDipoleMap}
\end{align}
and then map $\tilde{H}$ 
to  the $z$-plane. In the $z$-plane,
we work with the 
polar coordinate $(r,\theta)$ defined by
\begin{align}
x-r_0=:r_0\cos\theta,\quad 
y=:r_0\sin\theta.
\end{align}

By transforming $\tilde{H}$, 
the evolution operator in the $r$ direction is generated 
by\footnote{We get rid of the minus sign, considering the direction.}
\begin{align}
	\tilde{H}
	&=
	4r_0^3 \int^{2\pi}_0d\theta \cos^2 (\theta/2)\, T_{rr}(r_0,\theta).
\end{align}
Shifting the angular variable $\theta \equiv \phi+\pi$ for convenience, 
and introducing $L=  2 \pi r_0$ as well as
$\phi\equiv 2\pi s/L$, this reads
\begin{align}
	\tilde{H}
	&= 4r_0^3 \int^{2\pi}_0d\phi \sin^2 (\phi/2)\, T_{rr}(r_0,\theta)
	\nonumber \\
	&=   
	\frac{L^2}{\pi^2}
	\int^{L}_0ds\, \sin^2 \frac{\pi s}{L}\, 
	T_{rr}\left(\frac{L}{2\pi},\frac{2\pi s}{L}\right).
\end{align}
Thus, we have related $\tilde{H}$ to 
\begin{align}
	H_{\mathrm{SSD}}
	&= \int^{L}_0ds\, \sin^2 \frac{\pi s}{L}\, 
	T_{rr}\left(\frac{L}{2\pi},\frac{2\pi s}{L}\right).
\end{align}

By construction, we expect this SSD Hamiltonian has a CFT spectrum
on the infinite line,
although $H_{\mathrm{SSD}}$ 
is defined for a circle of circumference $L$.
The prefactor $L^2$ relating $\tilde{H}$ and $H_{\mathrm{SSD}}$
is indicative of the $1/L^2$ scaling. However,
since both, $H_{\mathrm{SSD}}$ 
as well as $\tilde{H}$  (defined on infinite space - see  Eq. (\ref{labelEquEvolutionDipoleMap}))
have a  continuum spectrum, there is no finite size scaling
that we can discuss for the level spacing.

\section{Inverse sine map}

\begin{figure}
\includegraphics[scale=0.35]{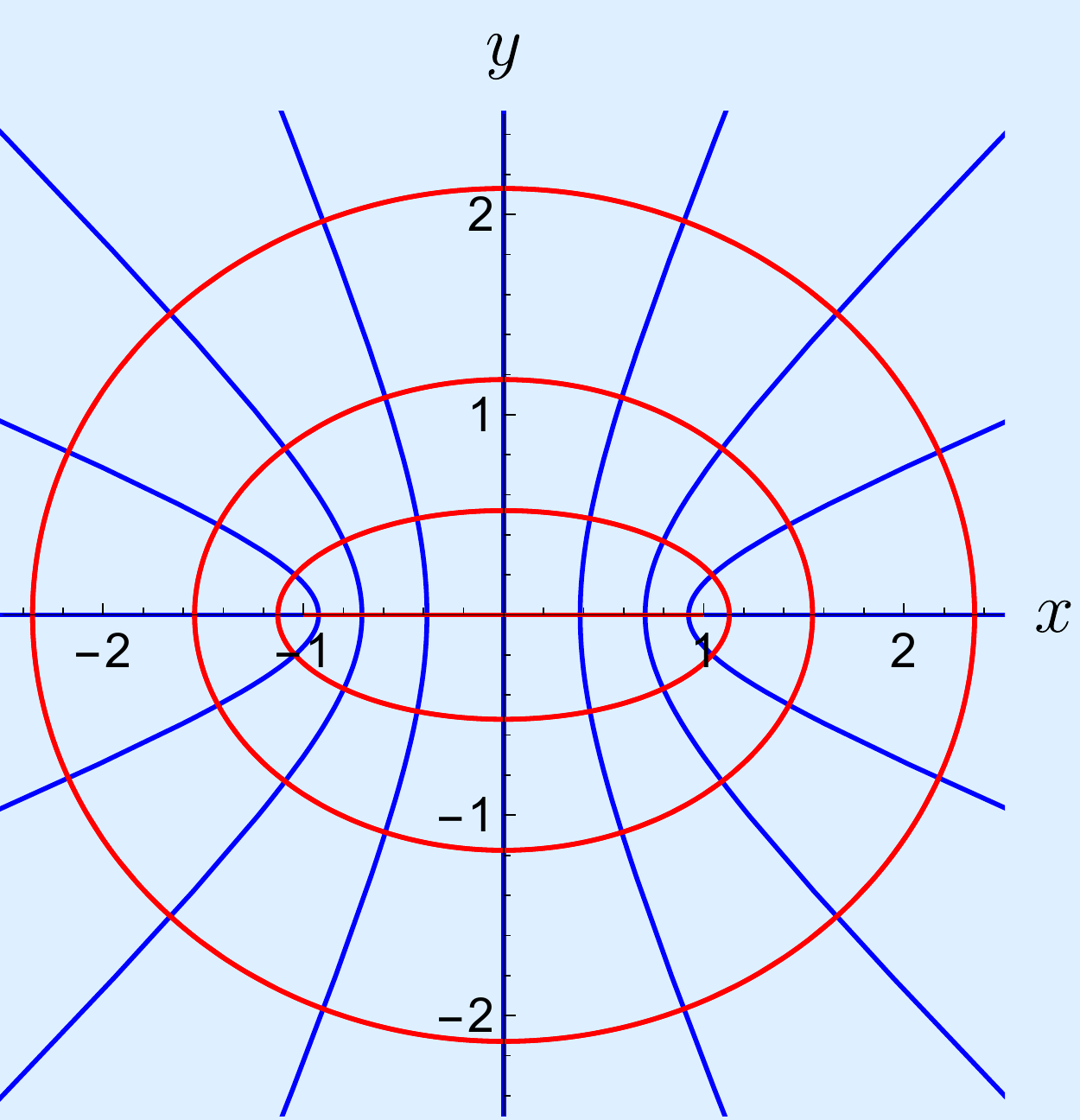} 
\includegraphics[scale=0.35]{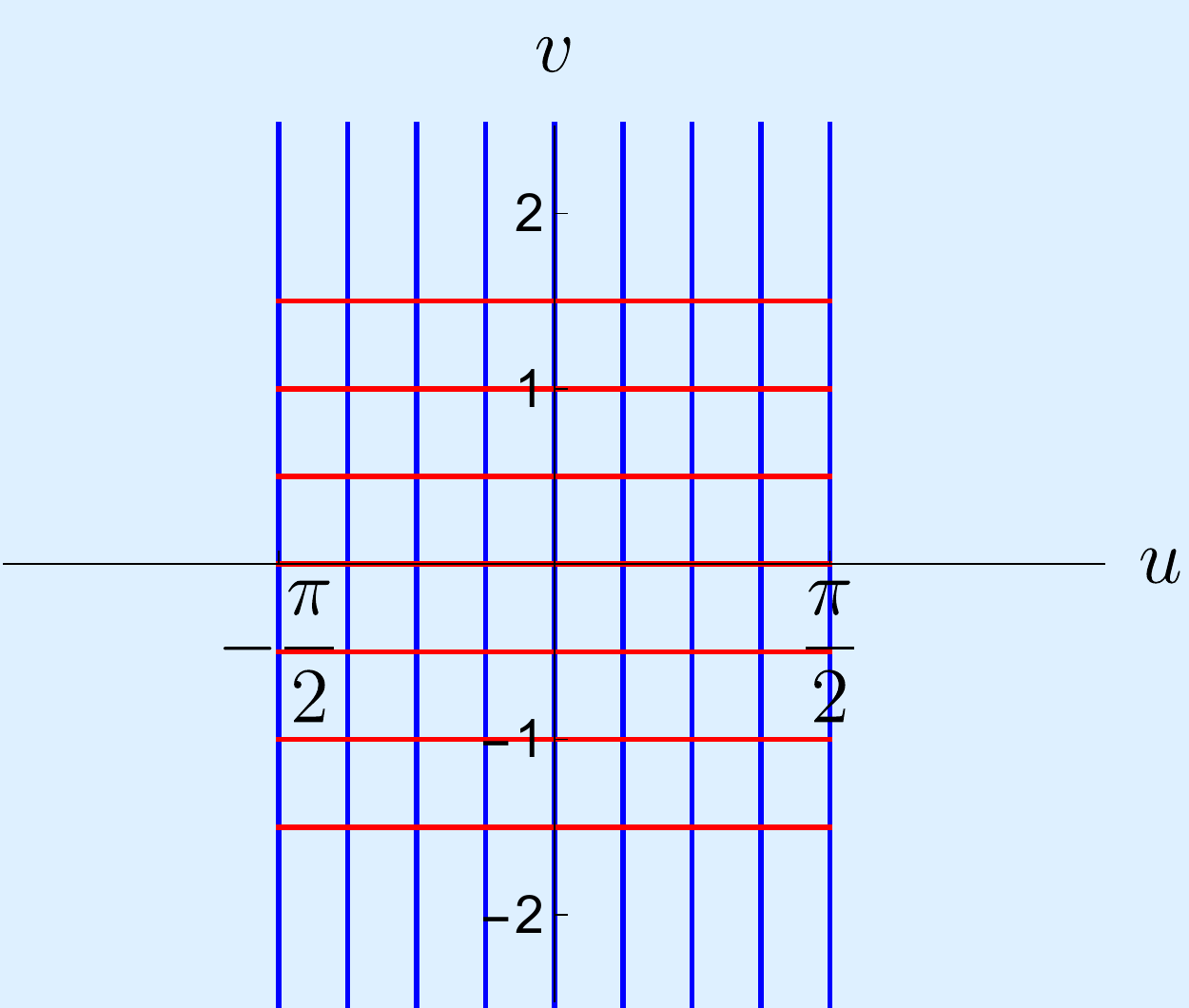}

\caption{Conformal map $z=\sin(w)$.}
\end{figure}

As yet another  conformal transformation, let us consider
\begin{align}
z=R\sin w. 
\end{align}
By this transformation, 
the infinite strip
defined by  $-\pi/2<u<\pi/2$ and $-\infty <v<+\infty$
is mapped onto the complex $z$-plane. 
%
%
%


Consider the 
evolution operator in the $v$-direction. At ``time'' $v=v_0$ this reads
\begin{align}
	\tilde{H}=\int^{+\pi/2-\delta}_{-\pi/2+\delta} du\, \tilde{T}_{vv}(u, v_0),
\qquad (\delta \to 0^+).
	\label{SRD start}
\end{align}
In the following, we focus on $y=0$, which translates into $v_0=0$.
Then, $x$ and $u$ are related by $x/R=\sin u$.

In \eqref{SRD start}, 
the upper and lower limit of the integral should be suitably cut off. 
As $u$ ranges over the interval  $(-\pi/2, +\pi/2)$, 
$x$ ranges over the interval  $(-R, R)$. 
When $x$ is close to its upper limit, $x=R-\epsilon$ where $\epsilon\to 0^+$, 
which suggests
\begin{align}
R-\epsilon=R\sin(\pi/2-\delta)=R(1-\delta^2/2+ ...),
\end{align}
and hence $\delta\sim\sqrt{{2 \epsilon}/R}\to 0^+$. 
In terms of $\epsilon$,  
the evolution operator is then written as
\begin{align}
	\tilde{H}=
	\int_{-\pi/2+\sqrt{2\epsilon/R}}^{+\pi/2-\sqrt{{2\epsilon/R}}}du\,
	\tilde{T}_{vv}(u, v=0). 
\end{align}

By mapping $\tilde{T}_{vv}(u,v)$ to $T_{yy}(x,y)$ we obtain
\begin{align}
	\tilde{H}=
	\int_{-R+\epsilon}^{+R-\epsilon}dx\,
	\sqrt{{R^{2}-x^{2}}}\, T_{yy}(x,y=0).
	\label{SRD cont}
\end{align}
We call this evolution operator 
"the square root deformation" (SRD).
This evolution operator is somewhat similar 
to the entanglement Hamiltonian.
However, this evolution brings (eventually, at asymptotically long times) the interval $x\in(-R,R)$
to to infinite space $x\in (-\infty,+\infty)$, unlike the
entanglement Hamiltonian which takes the interval $(-R,R)$ to its compliment on the real axis. 
Note that, interestingly, by construction the spectrum of 
the square root deformation, Eq. (\ref{SRD cont}),  does not scale with $R$.
%
%
%
%
%
%

\section{Numerics}

In this section, we use specific lattice models, such as 
the s=1/2 XX quantum spin model,
to study the deformed Hamiltonians.
We have also checked numerically the transverse Ising model and the XXZ model,
but the results for these models are qualitatively similar. 
We therefore focus here on the XX model.

\begin{figure}[t]
\includegraphics[scale=0.5]{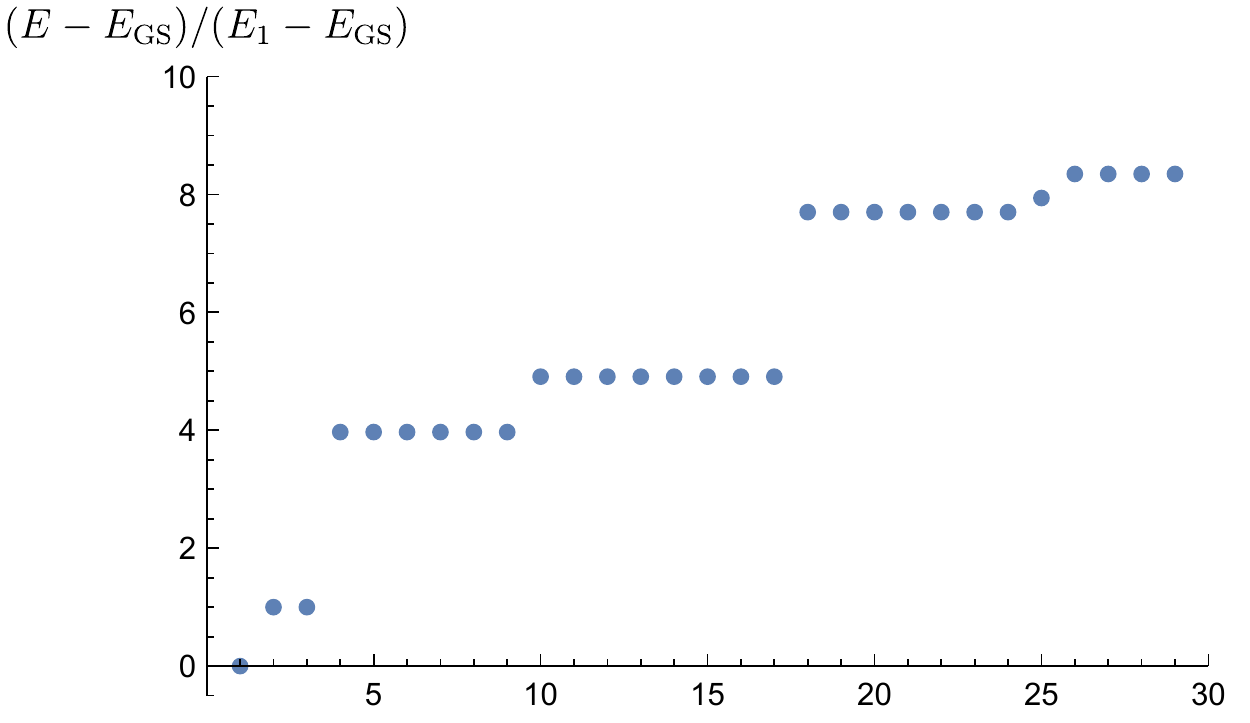}
\includegraphics[scale=0.5]{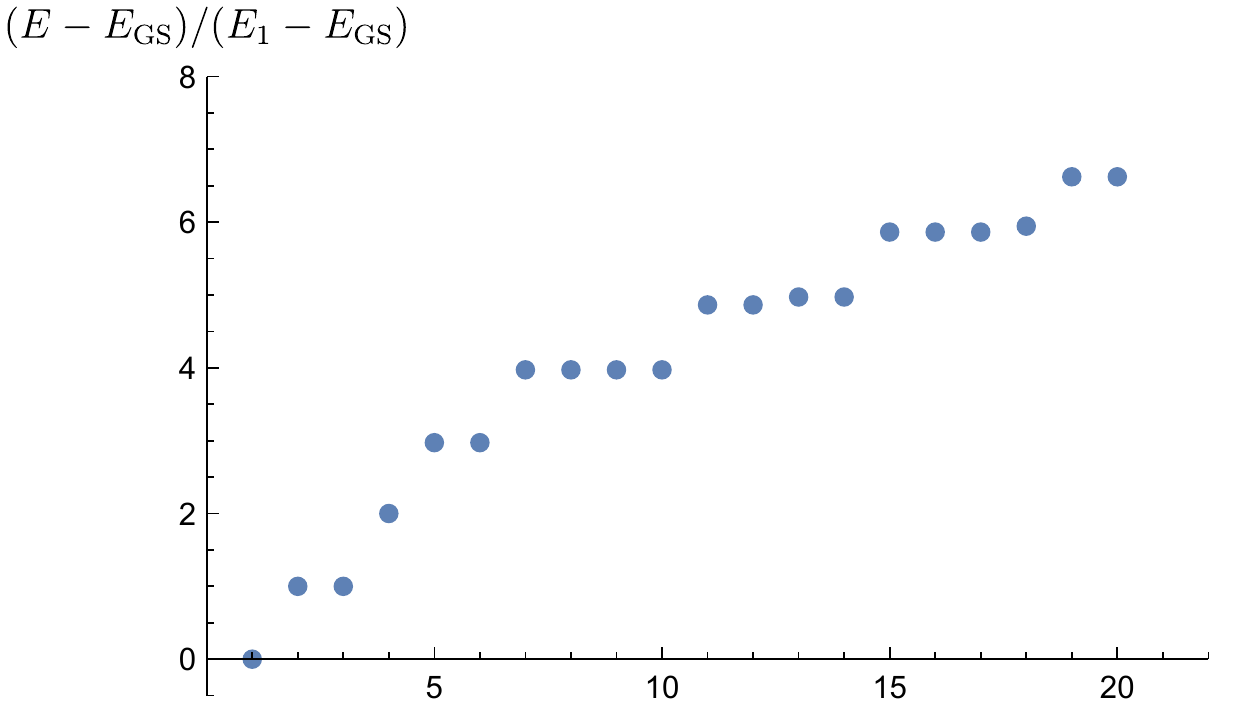}
\caption{
\label{XX PBC and OBC}
The finite size spectra of the XX model (18 sites) with PBC and OBC.}
\end{figure}

The spin 1/2 XX model is defined by:
\begin{align}
H=\sum_{i}\left(S_{i}^{x}S_{i+1}^{x}+S_{i}^{y}S_{i+1}^{y}\right)=\sum_{i}h_{i,i+1},
\end{align}
where $S^{x,y,z}_i$ is the spin 1/2 operator defined at site $i$ of a
one-dimensional lattice. 
The finite-size spectra of the XX model, both for PBC and for open boundary conditions (OBC), 
are shown in Fig. \ \ref{XX PBC and OBC}.
The low-energy part of these spectra are described by the $c=1$ compactified free boson theory. 
We will use these spectra as our reference when discussing the spectra of the 
deformed evolution operators. 

\subsection{SSD}

Let us now consider the deformation of the XX model by the envelope function:
\begin{align}
f(x)=\cos\frac{{2\pi x}}{L}+1
\end{align}
The resulting SSD Hamiltonian is given by
\begin{align}
H_{\mathrm{{SSD}}}=\sum_{i=1}^{L}f(x_{i}+1/2)h_{i,i+1}
\end{align}
where we impose the PBC, $h_{L,L+1}=h_{L,1}$. 
For previous analytical studies of the SSD of the XX model, 
see Ref.\ \onlinecite{2015JPhA...48R5208O}.

\begin{figure}[t]
\includegraphics[scale=0.5]{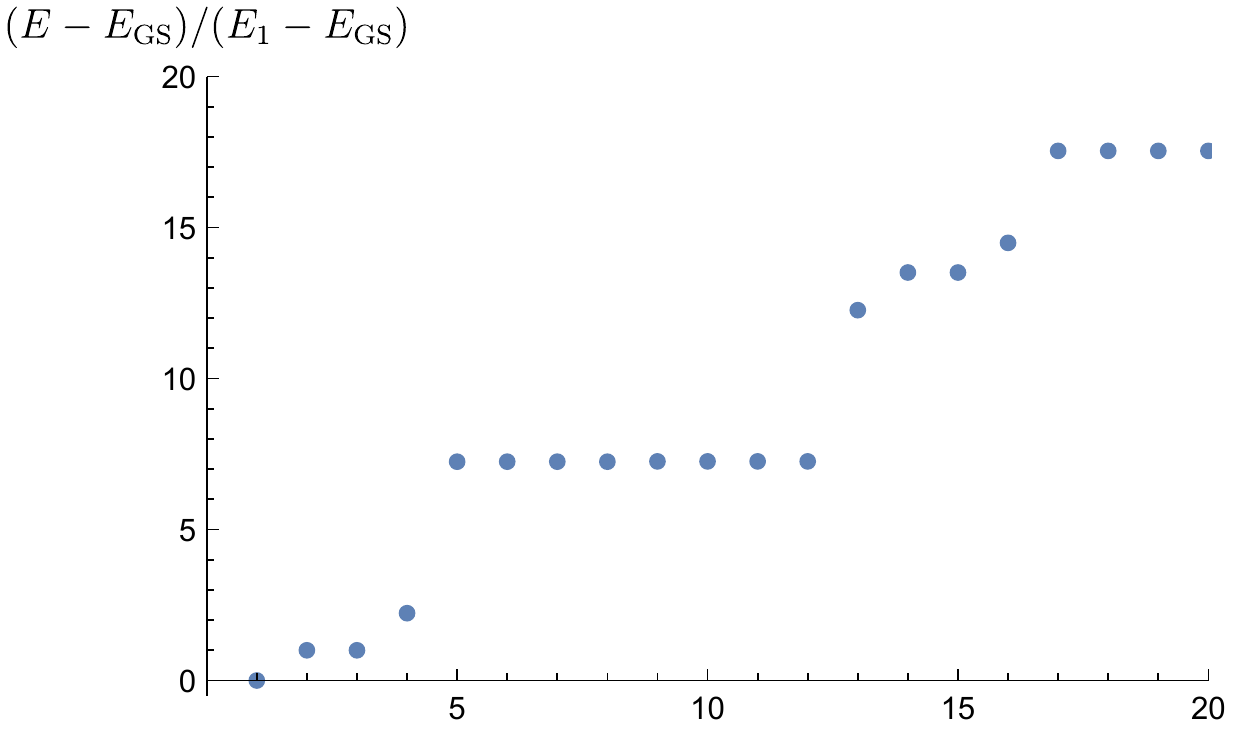}
\includegraphics[scale=0.5]{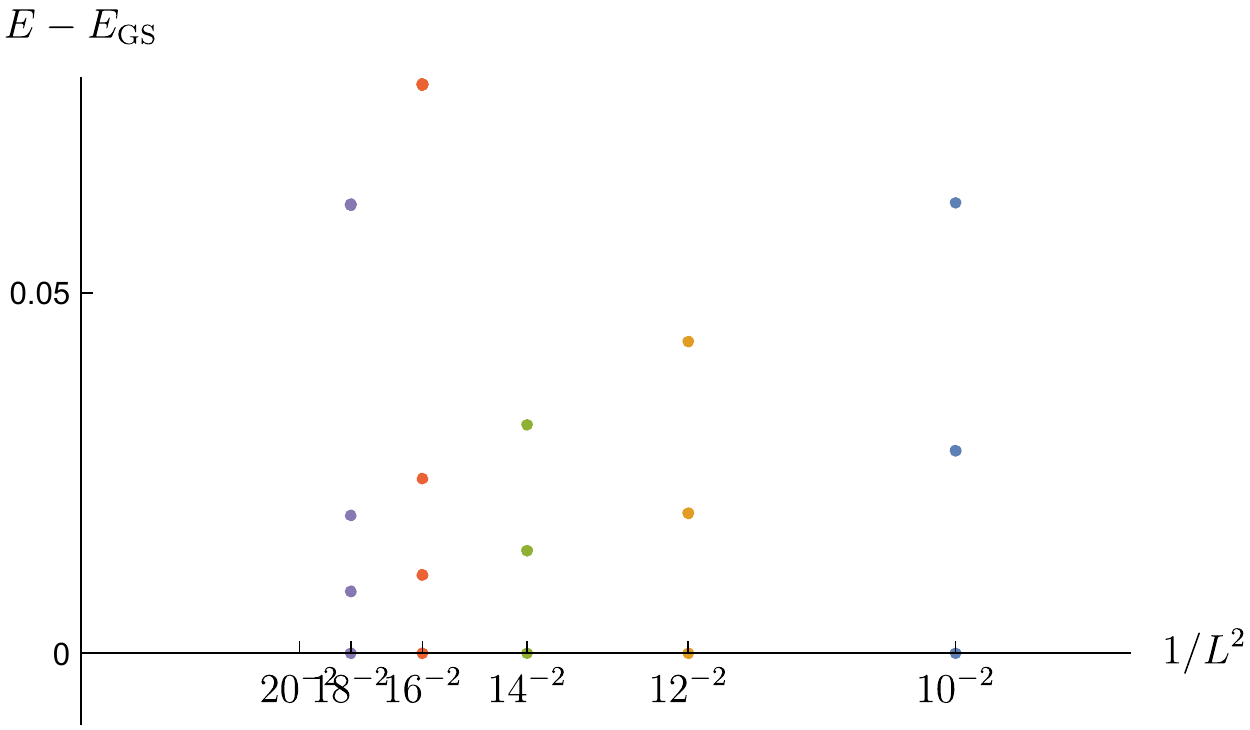}
\caption{
\label{SSD XX}
The finite size spectra of the XX model with the SSD (18 sites) and
the finite size scaling analysis of its low-lying spectrum.}
\end{figure}

In the continuum Hamiltonian,
we expect that this model 
exhibits a continuum spectrum even when the
system is put on a circle of finite circumference. 
The numerical, exact-diagonalization spectrum of the model shows a spectrum which 
does not compare well 
with the CFT spectrum, neither with PBC nor with  OBC
(Fig.\ \ref{SSD XX}), at least for the system sizes we studied.  

On the other hand,  finite size scaling analysis
shows the level spacing scales as $\sim1/L^{2}$ (Fig.\ \ref{SSD XX}).
This finding agrees with a previous numerical study. 
\cite{PhysRevB.87.115128}

\subsection{Regularized SSD}

Next, we turn to the regularized SSD.
The regularized SSD deformation of the XX model is given by the
envelope function
\begin{align}
f(x)=\cos\frac{2\pi x}{L}+\sqrt{{1-R^{2}/L^{2}}}.
\end{align}
Here $R$ here is a parameter which serves as a regularization. 
The envelope function reduces to  the SSD envelope function by taking
the limit $R\to 0$. 
The resulting SSD Hamiltonian is given by
\begin{align}
H_{\mathrm{{rSSD}}}=\sum_{i=1}^{L}f(x_{i}+1/2)h_{i,i+1}
\end{align}
where we impose the PBC, $h_{L,L+1}=h_{L,1}$. 

\begin{figure}[t]
\includegraphics[scale=0.5]{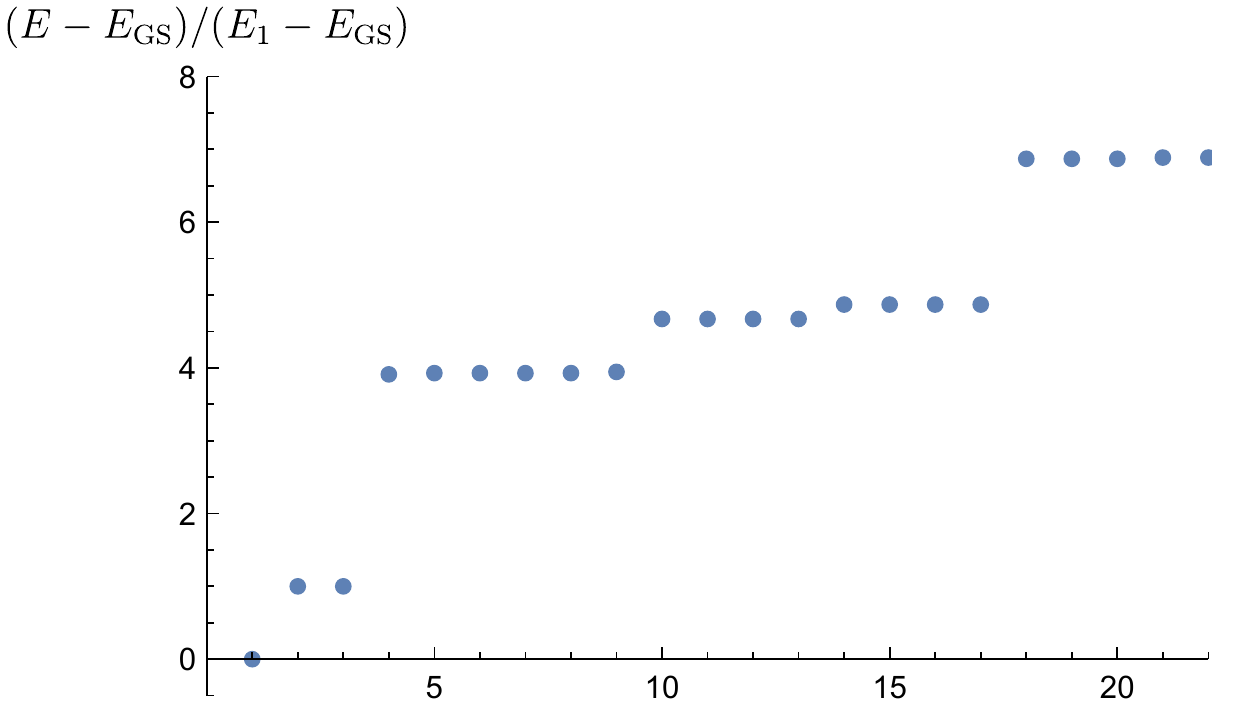}
\includegraphics[scale=0.5]{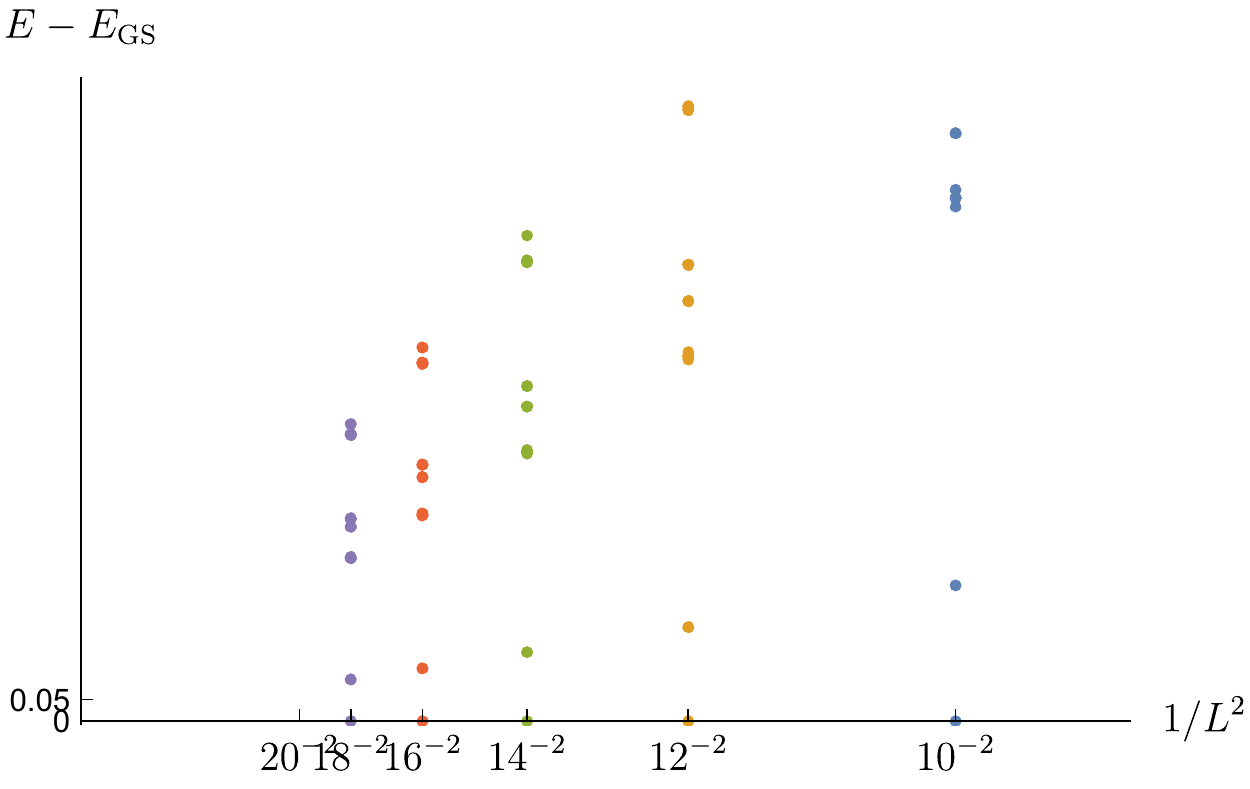}
\caption{
\label{RSSD XX}
The finite size spectra of the XX model with the regularized SSD (18
sites) and the finite size scaling analysis of its low-lying spectrum.
We chose $R=20.$}
\end{figure}

In the continuum limit,
we expect that this model exhibits the spectrum of a CFT defined on
a spatial circle. 
The low-lying part of the numerical exact-diagonalization
spectrum of the model 
(Fig.\ \ref{RSSD XX})
compares 
reasonably with the expected spectrum 
of the CFT with PBC in Fig.\ \ref{XX PBC and OBC}. 

As for the finite size scaling, we scale the system
size as well as the second term in the enveloping function. 
From the finite size scaling analysis within continuum field theories,
we expect the level spacing scales as $\sim1/L^{2}$. 
The numerical analysis in Fig.\ \ref{XX PBC and OBC} up to $L\sim20$ sites, 
where we choose $R=20$,
is in reasonable agreement with the expected $1/L^2$ scaling. 
On the other hand, 
we have checked that,
if we choose a smaller value of $R$, $R\sim1$, the low-lying spectrum
does not look like the CFT with PBC.

\subsection{Square root deformation (SRD)}

Finally, 
the square root deformation of the XX model is given by the envelope function
\begin{align}
f(x)
&=\sqrt{{(L/2)^2-(x-1/2-L/2)^{2}}}
\nonumber \\
&= \sqrt{ (x-1/2) (L-x+1/2) }. 
\end{align}
Observe that we have a "shift" by 1/2,
which ensures $f(L+1/2)=0$. 
The resulting SRD Hamiltonian is given by
\begin{align}
H_{\mathrm{{SRD}}}=\sum_{i=1}^{L}f(x_{i}+1/2)h_{i,i+1}
\label{SRD XX on lattice}
\end{align}
where $f(L+1/2)h_{L,L+1}=0$. 

\begin{figure}[t]
\includegraphics[scale=0.5]{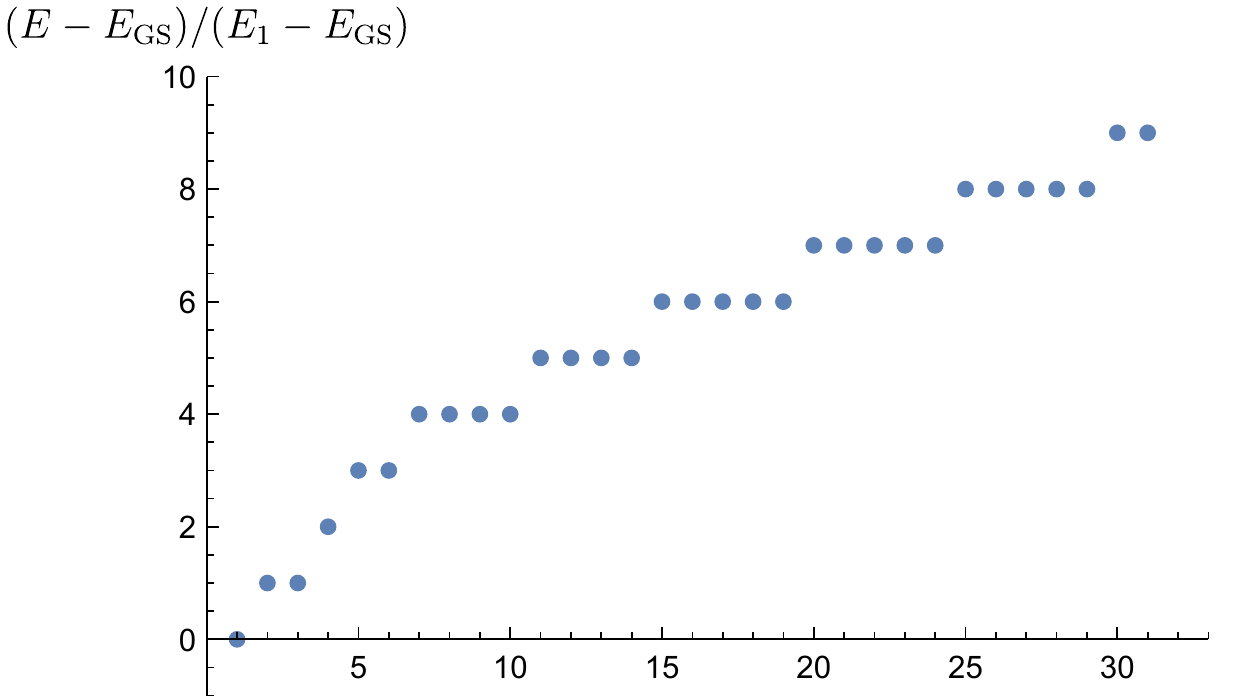}
\includegraphics[scale=0.5]{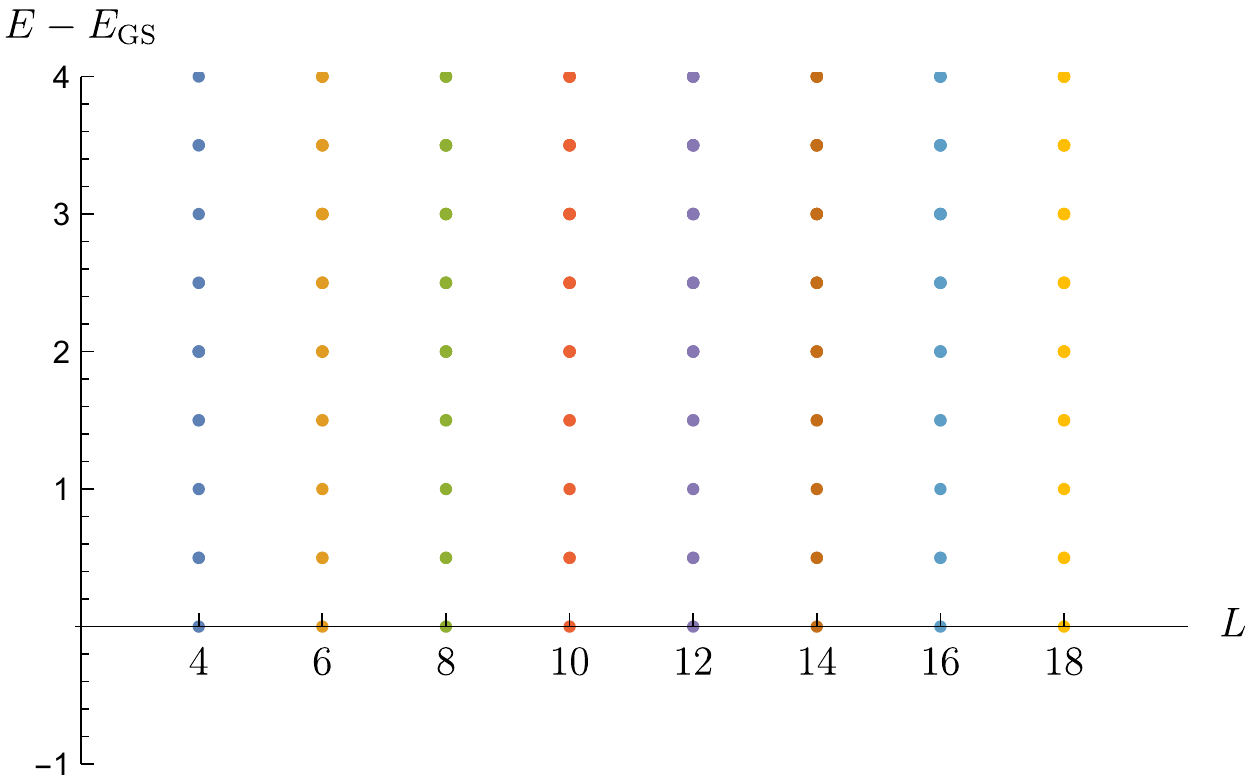}
\caption{
\label{SRD XX}
The finite size spectra of the XX model with the square root deformation
(18 sites) and the finite size scaling analysis of its low-lying spectrum.}
\end{figure}

In the continuum limit, we expect that this model exhibits the spectrum of a CFT with boundary. The low-lying
part of the numerical  exact-diagonalization spectrum of the model 
(Fig.\ \ref{SRD XX})
compares very well  
with the spectrum of the XX model with OBC 
(Fig.\ \ref{XX PBC and OBC}).
As for the level spacing, 
the finite size scaling analysis shows
the level spacing scales as $\sim1/L^{0}$ as expected. 
In fact, 
already from 
the very small system size ($L=4$),
the low-lying spectrum of 
$H_{\mathrm{SRD}}$ \eqref{SRD XX on lattice}
the finite size spectra of the XX model with OBC
agrees surprisingly well (almost perfectly).

In fact, this deformation is quite peculiar. When applied to a non-interacting
fermion system, it completely "straightens" out
the spectrum, and hence makes the spectrum perfectly relativistic,
as analytically shown in 
Ref.\ \onlinecite{2004PhRvL..92r7902C}.

%
%
%
%
%
%
%
\section{Conclusion}

Summarizing,
we have constructed various 
deformed evolution operators of type \eqref{H deformed}. 
In particular, we have constructed 
a regularized version of the SSD.
From our construction, it is also obvious that 
the regularized SSD Hamiltonian has a very close connection
with  the entanglement Hamiltonian;
the evolution generated by the regularized SSD and
the entanglement Hamiltonian are orthogonal to each other. 

We have also studied the scaling of the level spacing 
of the spectra of the deformed evolution operators. 
The regularized SSD shows  $1/L^2$ scaling
as opposed to 
(i) the regular $1/L$ scaling of CFTs put on a finite spatial circle of 
circumference $L$
and 
(ii) the scaling of the spectrum of the original SSD Hamiltonian.  
For the latter, within the continuum field theory description, 
the spectrum consists of a continuum (and hence there is no scaling
to be discussed for the level spacing).  
When the SSD Hamiltonians are studied on discrete one-dimensional lattices,
$1/L^2$ level spacing has been observed,
which seems closely related to the scaling of the regularized SSD. 

Generalization to  conformal maps
(e.g.\cite{2001JHEP...09..038R}, $z=(2/\pi)\mathrm{arctan}\, w$ )
other than  those we considered in this paper
should be straightforward.
To give a broader perspective, 
it is worth pointing out that our construction,
namely, the construction of the deformed evolution operator 
on the complex $z$-plane 
from some reference evolution operator on the $w$-plane (cylinder or strip),
is 
closely related to the classification scheme of 
conformal vacua
\cite{Birrell:1982ix,Candelas:1978gf} in curved spacetime.
In  
that  classification, 
we are interested in a curved spacetime $M$,
which is conformally mapped into the flat spacetime $\tilde{M}$.
We suppose 
that $\Sigma$, a global Cauchy hypersurface of $M$, 
is mapped under the conformal transformation 
to a global Cauchy hypersurface $\tilde{\Sigma}$ of $\tilde{M}$. 
Then, for a  timelike conformal Killing vector field  in $\tilde{M}$, 
there exists a global timelike conformal Killing vector field  in $M$. 
Thus, we can classify the conformal vacua
defined with respect to the latter conformal Killing vector field
by reference to the vacua defined in $\tilde{M}$. 
\cite{Birrell:1982ix}
In Ref.\ \onlinecite{Candelas:1978gf}, 
various (1+1) dimensional spacetimes are conformally mapped 
into the Einstein static universe (which can be represented as a spacetime cylinder.)

Taking the $z$-plane (the $w$-space) as $M$ ($\tilde{M}$), 
the only minor difference is that here we have considered
Euclidean conformal field theories,
while in Ref.\ \onlinecite{Candelas:1978gf}
conformal field theories in curved spacetime with  Minkowski signature 
are studied. 
At any rate, the global considerations as well as the classification scheme 
developed in Ref.\ \onlinecite{Candelas:1978gf}
can all  be 
applied to the set of deformed evolution operators 
discussed in this paper. 

{\it Note added:}
Upon completion of this work,
we became aware of the recent preprint Ref.\ \onlinecite{Okunishi},
in which the regularized SSD is also studied.

\acknowledgements

We thank Tom Faulkner and Hosho Katsura for useful discussion. 
We are grateful to the 
KITP Program ``Entanglement in Strongly-Correlated Quantum Matter''
(Apr 6 - Jul 2, 2015).
This work is supported by 
the NSF under Grants 
No.\ DMR-1455296 (X.W. and S.R.),
No.\ NSF PHY11-25915,
and 
No.\ DMR-1309667 (A.W.W.L.),
as well as 
by the  Alfred P. Sloan foundation. 

\bibliography{ref}

\end{document}